\documentclass[english,prl,aps,reprint,longbibliography]{revtex4-1}
\usepackage[latin9]{inputenc}
\usepackage{amsbsy}
\usepackage{amssymb}
\usepackage{graphicx}
\usepackage{esint}

\makeatletter

\newcommand*\LyXThinSpace{\,\hspace{0pt}}
\newcommand*\LyXHairSpace{\hspace{1pt}}
\DeclareTextSymbolDefault{\textquotedbl}{T1}

\makeatother

\usepackage{babel}
\begin{document}
\title{Half-Quantized Hall Effect and Power Law Decay of Edge Current Distribution}
\author{Jin-Yu Zou, Bo Fu, Huan-Wen Wang, Zi-Ang Hu, Shun-Qing Shen}
\email{sshen@hku.hk}

\affiliation{Department of Physics, The University of Hong Kong, Pokfulam Road,
Hong Kong, China}
\date{\today}
\begin{abstract}
The half-quantized Hall conductance is characteristic of quantum systems
with parity anomaly. Here we investigate topological and transport
properties of a class of parity anomalous semimetals, in which massive
Dirac fermions coexist with massless Dirac fermions in momentum space
or real space, and uncovered a distinct bulk-edge correspondence that
the half-quantized Hall effect is realized via the bulk massless Dirac
fermions while the nontrivial Berry curvature is provided by the massive
Dirac fermions. The spatial distribution of the edge current decays
away from the boundary in a power law instead of an exponential law
in integer quantum Hall effect. We further address physical relevance
of parity anomalous semimetal to three-dimensional semi-magnetic topological
insulators and two-dimensional photonic crystals.
\end{abstract}
\maketitle

\paragraph{Introduction}

The massless Dirac fermion in 2+1 dimensions coupled with an electromagnetic
field $A_{\mu}$ is invariant under time-reversal and spatial reflection
symmetry at the classical level but loses the symmetries when the
theory is quantized in a gauge-invariant fashion. This phenomenon
is known as ``parity anomaly'' \citep{Niemi1983axial,Jackiw1984Fractiona,Redlich1984gauge,Boyanovsky1986physical,Schakel1991relativistic}.
Upon integrating out the fermion field in the action by means of the
Pauli--Villars regularization \citep{Redlich1984gauge}, a Chern-Simons
term which is odd under reflection and time reversal arises in the
effective Lagrangian for the gauge field: $\mathcal{L}_{\mathrm{CS}}[A_{\mu}]=\frac{\sigma_{H}}{2}\epsilon^{\mu\nu\rho}A_{\mu}\partial_{\nu}A_{\rho}$
with $\epsilon^{\mu\nu\rho}$ the Levi-Civita symbol and $\sigma_{H}=\frac{1}{2}\frac{e^{2}}{h}\frac{M}{|M|}$
only depending on the sign of the mass $M$ for the regulator. The
Chern-Simons term predicts a half-quantized Hall conductance \citep{Qi2008topological}.
Several condensed matter systems were proposed to realize parity anomaly
in early pioneering works, such as the single layer graphite system
\citep{SemenoffPRL1984}, PbTe-type narrow gap semiconductor with
a domain wall \citep{FradkinPRL1986} and recently in HgTe heterostructure
\citep{B=0000F6ttcher2019survival}. Haldane proposed that the half-quantized
Hall conductance could be obtained if one of the two gaped valleys
on a honey comb lattice can be fine-tuned to be closed \citep{Haldane1988Model}.
Recently, the Haldane's idea has been constructed in photonic \citep{Liu2020NC}
and phonon \citep{Lu2017NP,Ding2019PRL} crystal, as well as Floquet
systems \citep{Hu2020npj,Leykam2016prl}. Another attempt to realize
the parity anomaly is based on the surface states of three-dimensional
(3D) $Z_{2}$ topological insulators \citep{Fu2007topological,Qi2008topological,hasan2010colloquium,qi2011titsc,shen2012topological}.
If the surface states are gapped by magnetic doping or proximity effect
at one surface of the system while the surface states at the opposite
surface remain gapless, an unpaired gapless Dirac cone can be realized
in the quasi-2D system with a lattice regularized description \citep{Lee2009surface,Chu-11prb,Rosenberg2010Wormhole,Nomura2011surface,Brune2011quantum,Mulligan2013topological,Koenig2014half,Zhang2017anomalous,Lapa2019parity,Burkov2019Dirac}.
Recently, this 'semi-magnetic' heterostructure was reported experimentally
in Ref. \citep{Lu2021half,Mogi2021arxiv}. As illustrated in Fig.
1(a) and (b), the massive and massless Dirac cones are separated in
momentum space in Haldane model and in position space in semi-magnetic
topological insulator, which plays the role of the Pauli-Villars regulator,
making them as an ideal platform for the condensed matter realization
of the ``parity anomaly''. The coexistence of massive and massless
Dirac fermions constitute a topological semimetal, named ``parity
anomalous semimetals'', to emphasize the role of the massless Dirac
fermions for parity anomaly.

Opposite to the integer quantum Hall effect and quantum anomalous
Hall effect \citep{Thouless1982quantized,HatsugaiPRL1993}, the half-quantized
Hall effect does not indicate the existence of well defined chiral
edge states as the number of chiral edge state cannot be half of an
integer, and also the carrier charge is not fractional of elementary
charge $e$ in these non-interacting systems. In this Letter, we study
electronic and transport properties of parity anomalous semimetal
for an open boundary condition by means of a combination of analytical
and numerical techniques. While no well-defined edge states are present,
it is found that the accumulation of all extended bulk states exhibit
chiral nature and an edge current circulates around the boundary as
shown in Fig. 1(c). The edge current decays in a power law behavior
of the distance $x$ away from the boundary $x^{-3/2}$, which is
different from the exponential decay in integer quantum Hall effect
due to the presence of the localized edge states \citep{shen2012topological,MongPRB2011,VandenbergheNC2017}.
The circulating electric current generates a magnetization which leads
to a half-quantized Hall conductance according to Streda formula \citep{Streda1982,Sinitsyn2007}.
We also demonstrate the topological charge pumping from the collective
contribution of the bulk extended states. Furthermore, we explore
the Hall current distribution and charge pumping in a semi-magnetic
topological insulator slab, which provides a comprehensive and distinct
physical picture for the half-quantized surface Hall effect.

\begin{figure}
\begin{centering}
\includegraphics[width=8cm]{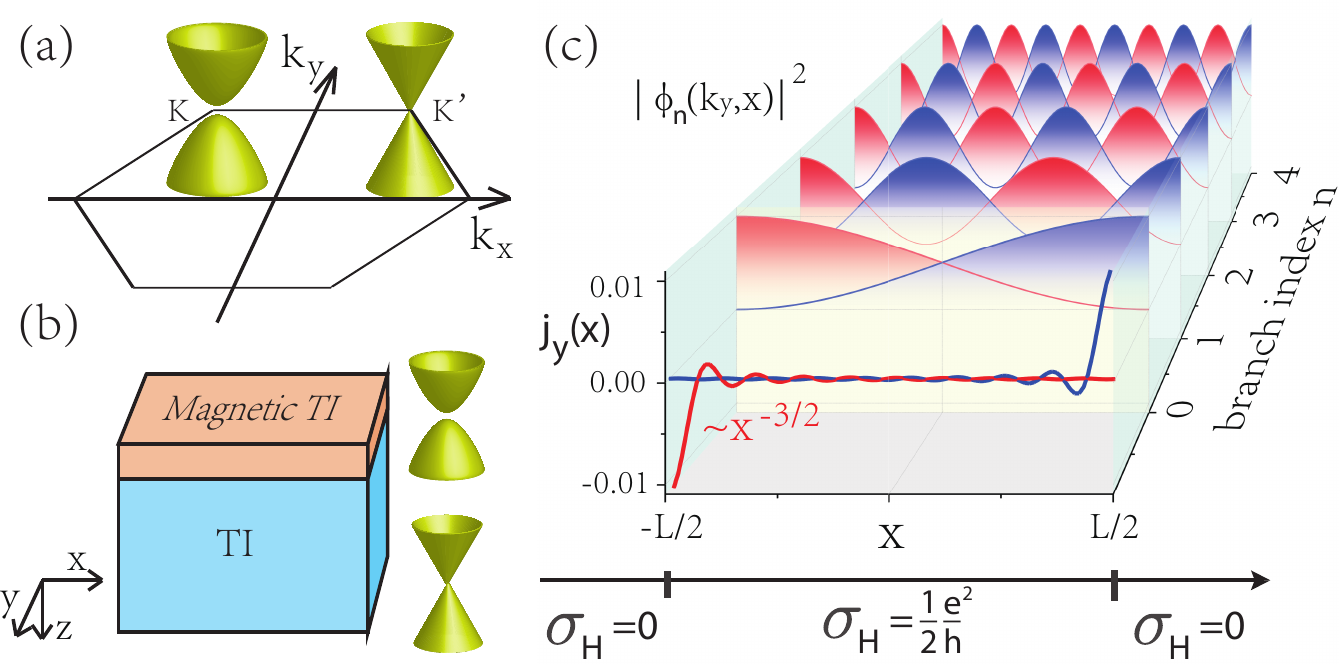}
\par\end{centering}
\caption{Illustration of parity anomaly semimetals: (a) the Haldane model where
massive and massless Dirac cone separated in momentum space and (b)
the semi-magnetic 3D topological insulator in which a massive and
a massless Dirac cone separated in position space. (c) The distribution
of a set of low energy states and the power law decay edge current
in the parity anomalous semimetal with the half quantized Hall conductance
$\sigma_{H}=\frac{1}{2}\frac{e^{2}}{h}$ for open boundary condition.
The up- and down- propagating states are indicated by red and blue
colors, respectively.}
\end{figure}

\paragraph{Haldane model with parity anomaly}

To demonstrate the distinct bulk-edge correspondence, we consider
the Haldane model with armchair-type termination as illustrated in
Fig. 2(a). To meet the boundary condition of the armchair termination,
we must admix states of different valleys\citep{Brey2006electronic}.
Effectively, the presence of the gapped valley behaves as a boundary
conditions for the envelope function of the gapless valley\citep{Note-on-SM}.
If we consider the low energy scale which is much smaller than the
gap of the massive valley, the confinement can be approximated as
the hard-wall boundary condition which was extensively discussed in
previous studies \citep{Berry1987,McCann2004,Tworzyd=000142o2006subpoissonian,Ferreira2013magnetically,Resende2017confinement}.
In this approach, the energy eigenvalues are analytically solvable
$\varepsilon_{n}(k_{y})=\hbar v_{F}\sqrt{k_{y}^{2}+[\frac{\pi}{L}(n+\frac{1}{2})]^{2}}$
($n=0,1,2,3...$) with $v_{F}$ the fermi velocity and $L$ being
the length of the ribbon. Due to the presence of factor $1/2$, there
exists a finite energy gap inversely proportional to the ribbon width
$L$, $\Delta=2\hbar v_{F}\pi/L$. The numerical eigenvalues for the
eigenenergy shows good agreement with this analytic expression in
the low energy scale. The corresponding wave function with conserved
momentum $k_{y}$ can be written as $\Psi_{n}(x,y)=e^{ik_{y}y}|nk_{y}\rangle$
which are all extended states. However, we can calculate the average
displacement relative to the center of the ribbon for each states
$\langle nk_{y}|(\frac{x}{L}-\frac{1}{2})|nk_{y}\rangle=-\frac{\hbar v_{F}k_{y}}{\varepsilon_{n}(k_{y})}\frac{2}{\pi^{2}(2n+1)^{2}}$
which is proportional to the propagation velocity. This result means
that the down propagating states locates in the left half portion
of the ribbon and the states locate in the opposite portion flow in
the opposite directions (shown by the red and blue lines with filling
color in Fig. 1(c), respectively). To indicate this chiral nature,
we calculate the local density of states at the boundary in Fig. 2(b).
There is a remarkable asymmetry in the spectral weight between the
left-and right-moving modes. As a consequence, there must be an equilibrium
circulating current \citep{B=0000FCttiker2009edge} in the sample
for finite chemical potential. The transverse current density is given
by $j_{y}(x)=e\sum_{k_{y}}^{\varepsilon_{n}(k_{y})<\mu}\langle nk_{y}|\frac{1}{\hbar}\frac{\partial H}{\partial k_{y}}|nk_{y}\rangle$
for the occupied states\citep{Note-on-SM}
\begin{equation}
j_{y}(x)=\frac{e}{h}\frac{\mu}{2}\sum_{s=\pm}s\frac{J_{1}(2k_{F}|x-R_{s}|)}{|x-R_{s}|}\label{eq:current}
\end{equation}
where $\mu$ is the chemical potential, $k_{F}=\mu/(\hbar v_{F})$
is the Fermi wave vector, $J_{n}(x)$ is the first kind Bessel function
and $R_{s=+}=0$ or $R_{s=-}=L$ for two edges $s=\pm$. By using
the asymptotic form of $J_{1}(x)$ at a large argument $x$, we find
the edge current $j_{y}^{s}(x)\propto x^{-3/2}\cos(2k_{F}x-\frac{3\pi}{4})$
maximizes at the boundary and decays to zero in a power law$\sim x^{-3/2}$
when moving away from the boundary with the oscillation length as
$1/(2k_{F})$. This analytic result fits well with the numerical calculation
as shown in Fig. 2(c). As $\mu$ increases, the oscillation length
of $j_{y}^{s}(x)$ decreases. In a quantum anomalous Hall insulator
or integer quantum Hall insulator, the localized edge states give
an exponential decay behavior with the decaying length proportional
to the inverse of the energy gap. This power law behavior of the equilibrium
circulating current with the decay exponent $-3/2$ provides a fingerprint
to identify the parity anomalous semimetal phase. The magnetization
$\boldsymbol{M}$ induces a magnetization current density $\mathbf{\boldsymbol{j}}=\nabla\times\boldsymbol{M}$.
Inversely, the equilibrium current $j_{y}^{s}(x)$ corresponds to
a spatially varied magnetization
\begin{equation}
M_{z}^{s}(x)=-\int_{R_{+}}^{x}dx^{\prime}j_{y}^{s}(x^{\prime})=-\frac{e}{h}\frac{\mu}{2}\mathcal{F}(2k_{F}(x-R_{+}))\label{eq:magnetization}
\end{equation}
with $\mathcal{F}(x)=J_{1}^{\prime}\left(x\right)+\frac{\pi x}{2}\left[J_{1}(x)H_{0}(x)-J_{0}(x)H_{1}(x)\right]$
where $H_{n}(x)$ are the Struve functions. $\mathcal{F}(x)$ has
the asymptotic behaviors $\mathcal{F}(x)=x/2$ for a tiny $x\rightarrow0$
and $\mathcal{F}(x)=1$ for $x\rightarrow+\infty$. The magnetic field
$B_{z}=\mu_{0}M_{z}$ through the sample as a function of position
which should be measurable by magnetic flux detectors such as the
superconducting quantum interference device (SQUID) \citep{Uri2020nanoscale}.
We plot the magnetization field $M_{z}=\sum_{s=\pm}M_{z}^{s}$ versus
position $x$ in armchair ribbon by setting the origin at one edge
shown as the yellow curve in Fig. 2(c). As moving from the edge, the
magnetization $M_{z}$ increases from $0$ and then saturates to the
bulk value with oscillation. Due to finite size confinement along
$x$ direction, the magnetization is not half-quantized (in the unit
of $e\mu/h$) when the ribbon is narrow. However, the bulk value of
$M_{z}$ converges quickly into the half quantized value as $k_{F}L$
increases owing to the asymptotic behavior of $\mathcal{F}(x)$. The
total equilibrium out-of-plane magnetization can be calculated thermodynamically,
$M_{z}=-\partial_{B}\Omega(\mu,B)$ where $B$ is the magnetic field
and $\Omega$ is the grand-canonical thermodynamic potential of electrons
at chemical potential $\mu$. From the Maxwell relation, we have $\partial_{\mu}M_{z}=-\partial_{\mu}(\partial_{B}\Omega)=\partial_{B}(-\partial_{\mu}\Omega)=\partial_{B}\rho$
with $\rho$ the carrier density. In the thermodynamic limit $k_{F}L\to\infty$,
the Hall response $\sigma_{H}=e\partial_{\mu}M_{z}=\frac{e^{2}}{2h}$
is half-quantized according the Streda formula \citep{Streda1982}
$\sigma_{H}=e\partial_{B}\rho|_{\mu}$. The bulk magnetization can
be obtained from the edge current $M_{z}^{bulk}=J_{\mathrm{edge}}=\int_{0}^{L/2}dxj_{y}(x).$
As shown in Fig. 2(d), we plot the edge current $J_{\mathrm{edge}}$
as a function of the chemical potential in the armchair ribbon (blue
line with circles). When the chemical potential is around the band
crossing point, the edge current displays a linear dependence of $\mu$
with a slope of $\sim e/2h$.

\begin{figure}
\begin{centering}
\includegraphics[width=8cm]{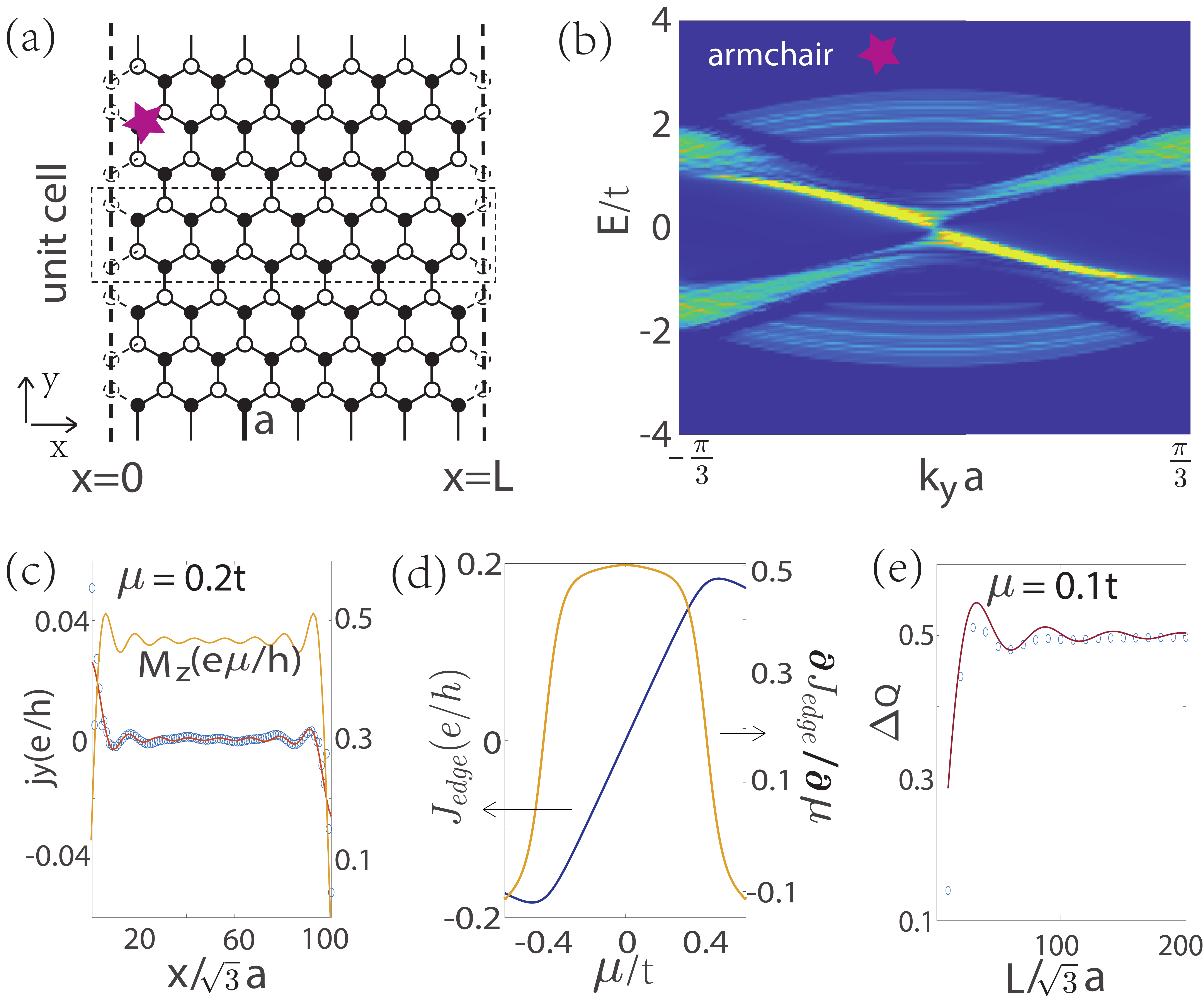}
\par\end{centering}
\caption{(a) Structure of honeycomb lattice nanoribbons with armchair edges.
$a$ is the lattice distance. (b) The local density of states at the
left edge (labeled by the star). The yellow color indicates higher
values and blue indicates the lower values. The dots and lines are
numerical and analytical results, respectively. (c) The distribution
of current $j_{z}$ and magnetization $M_{z}$ with the Fermi level
$\mu=0.2t$. (d) The edge current (blue line) $J_{\mathrm{edge}}=\int_{0}^{L/2}dxj_{y}(x)$
and its derivative with respect to $\mu$ (orange line) as a function
of $\mu$.(e) The charge pumping $\Delta Q$ as a function of ribbon
width $L$ for a fixed $\mu$. The dots and line are numerical and
analytical (Eq. (\ref{eq:charge_transfer})) results, respectively.}
\end{figure}

\paragraph*{Charge pumping}

We consider a cylinder with a circumference $W$ along the $y$ direction
and with length $L$ along the $x$ direction , which is pierced by
magnetic flux $\Phi$. With changing the magnetic field, an electric
field along the circumferential direction can be induced according
to the Faraday's law, $E(t)=\frac{1}{W}\partial_{t}\Phi$. Due to
boundary confinement effect along $x$ direction, the energy spectrum
for such geometry becomes a series discrete one-dimensional subbands
$\varepsilon_{n}$. In the presence of the electric field, the acceleration
of the electron is given by $\hbar\dot{k}=eE$. The magnetic flux
is switched on adiabatically which means that the rate of change of
the flux is much smaller than the energy spacing between two bands
$\hbar v_{F}\pi/L$, such that any particles are not excited into
the next subbands. Then electrons with positive velocity increase
while electrons with negative velocity decrease while the total number
of electrons for each branch $n$ is conserved. Over the interval
time $\Delta t$, the change of the wavevector is $\frac{e}{\hbar}E\Delta t$.
If the flux changes by one quantum flux $\Phi_{0}=h/e$ over the time
$\Delta t$, we have $EW\Delta t=\Phi_{0}$. The spatial imbalance
for each state is $\Delta\rho_{n}(k_{x})=\langle nk_{y}|[\Theta(x)-\frac{1}{2}]|nk_{y}\rangle=\frac{\hbar v_{F}k_{y}}{\varepsilon_{n}(k_{y})}\frac{\sin(\pi(n+1/2))}{2\pi(n+1/2)}$,
where $\Theta(x)$ is the step function. Thus the charge transfer
from one side to the other can be expressed as the spatial imbalance
difference between the two states with opposite velocity at the fermi
level $\Delta Q=\frac{e}{2}\sum_{n}\left[\Delta\rho_{n}(k_{n}^{f+})-\Delta\rho_{n}(k_{n}^{f-})\right]$\citep{Note-on-SM},
where we have assumed the chemical potential $\mu$ lies in the gap
of the gapped valley and intersects only with the gapless one and
$k_{n}^{f\pm}$ are the two Fermi wave vectors for branch $n$. The
summation over $n$ is performed over all the branches intersecting
the Fermi level and can be done analytically, 
\begin{equation}
\Delta Q=-\frac{e}{2}\mathcal{F}(k_{F}L).\label{eq:charge_transfer}
\end{equation}
 $\mathcal{F}(x)=1$ for $x\rightarrow+\infty$. As shown in Fig.
2(e), we plot the analytical expression (Eq. (\ref{eq:charge_transfer}))
and the numerical results for the cylinder geometry as a comparison.
Thus there is half charge pumping from the one edge to the other in
in the thermodynamic limit. The final expression for charge transfer
(\ref{eq:charge_transfer}) is a collective consequence from all the
bands intersecting the Fermi level. It is in sharp contrast with quantum
anomalous Hall insulator where the charge transfer is attributed from
the two chiral edge states at the two sides \citep{Laughlin1981quantized,Halperin1982quantized,Thouless1983quantization,Niu1984quantised}.
Assuming that we have a system with the Hall conductance $\sigma_{H}$,
we obtain the charge transferred as $\Delta Q=\sigma_{H}\Phi_{0}$.
From Eq. (\ref{eq:charge_transfer}), the Hall conductance is given
by $\sigma_{H}=-\frac{e^{2}}{2h}\mathcal{F}(k_{F}L)$ which is identical
to the expression derived from the magnetization (\ref{eq:magnetization})
by setting $x$ at the center of the ribbon. It indicates that the
half charge pumping shares the same topological origin with the half
quantized quantum Hall effect.

For the zigzag boundary condition there exists the localized states
along the boundary and will slightly revise the picture \citep{Brey2006electronic}.
For the extended states, the zigzag boundary condition will not admix
two valleys. We can solve the enveloped functions for each valley
separately and find that the spatial imbalance vanishes $\Delta\rho_{n}(k_{n}^{f\pm})=0$
for all extended states. As for the localized edge states which connect
two valleys, the boundary condition will strongly admix valley states.
The localized states solutions depend on the wave number and only
exist in lowest energy branch $n=0$. The positive Fermi vector corresponds
a localized state with $\Delta\rho_{n=0}(k_{n=0}^{f+})=1$ while the
negative Fermi vector corresponds an extended state with $\Delta\rho_{n=0}(k_{n=0}^{f-})=0$.
Consequently, we have $\Delta Q=\frac{e}{2}\Delta\rho_{n=0}(k_{n=0}^{f+})=\frac{e}{2}$
\citep{Note-on-SM}.

\begin{figure}
\begin{centering}
\includegraphics[width=8cm]{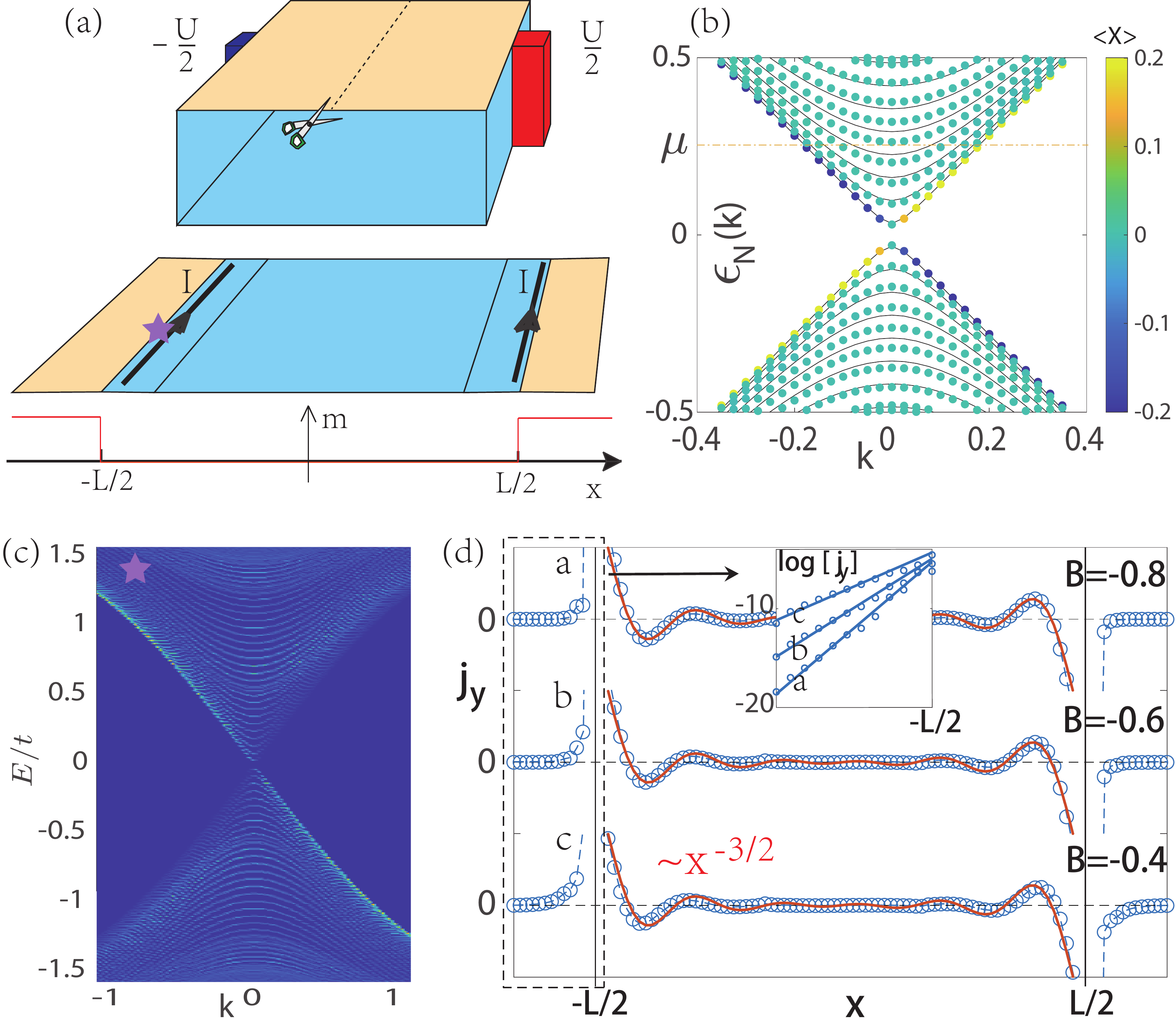}
\par\end{centering}
\caption{(a) A slab of three dimensional topological insulator coated by an
insulating ferromagnetic material (yellow region) on the top. When
the voltage bias $U$ is applied by the leads attached on the $x$
terminals, net current presents on the hinge. (b) The comparison of
the numerical results of the spectrum of 3D semi-magnetic topological
insulator for quasi-1D geometry (dots) and the analytic expression
$\varepsilon_{n}(k)=\hbar v_{F}\sqrt{k^{2}+[\frac{\pi}{L}(n+\frac{1}{2})]^{2}}$
(line). The color represents the center of mass $\langle x/L\rangle$
for each state. (c) The local density of states at the top hinge at
the position labeled as star in (a). (d) The surface current distribution
on the perimeter of the semi-magnetic topological insulator bar for
a fixed chemical potential $\mu=0.15m$ and several different Zeeman
fields $B=-0.2m$, $-0.3m$ and $-0.4m$ induced by the coated ferromagnetic
material. Insets shows $\mathrm{log}(j_{y})-x$ plot in the gapped
region. The red lines are the analytic current distributions according
to Eq. (\ref{eq:current}). We have chosen the bulk band gap $m$
as the energy unit and $\hbar v_{F}=0.7am$ with $a$ the lattice
constant.}
\end{figure}

\paragraph{3D semi-magnetic topological insulator}

An another potential candidate of parity anomalous semimetal is the
3D topological insulator coated by an insulating ferromagnetic material
on its top as shown in Fig. 3(a), now named as semi-magnetic topological
insulator \citep{Mogi2021arxiv}. The topological insulator hosts
massless Dirac fermions around its surface. The states located at
the interface between the topological insulator and the ferromagnet
open an energy gap due to the proximity effect. Thus the massive Dirac
fermions and massive Dirac fermions are separated in space, but still
have to coexist to form a semimetal as a whole becausemassive Dirac
fermions alone are prohibited to exist independently. To illustrate
the topological properties of the system, we can image to unfold the
surfaces states into a flat 2D plane \citep{Hattori2016Chiral}: the
gapless central region $|x|<L/2$ is sandwiched between the two gapped
outer regions $|x|>L/2$. We plot the numerical results of the energy
spectrum for the 3D semi-magnetic topological insulator (colored dots)
as a comparison with the analytical expression (black lines) in Fig.
3(b). We denote the bulk band gap by $m$ and the surface band gap
induced by the Zeeman field by $B$ with $|B|<|m|$. Here we only
consider the physics in the energy window that the chemical potential
is located within the surface band gap $|B|$. The numerical and analytical
results are in good agreement with each other in this low energy region.
Since the top and bottom surface Dirac cone are located at the same
point in momentum space, its spectra and physical properties are analogue
to the critical Haldane model with the armchair termination.

Due to the quasi-2D feature of semi-magnetic topological insulator,
it will display some unique features distinct from the 2D system.
One can also calculate the Hall conductance explicitly as a sum of
the real space projected layer-resolved Hall conductance $C_{z}(l)$,
$\sigma_{\mathrm{H}}=\sum_{l}C_{z}(l)e^{2}/h$ \citep{Essin2009PRL,sitte2012topological,wang2015Quantized,Note-on-SM}.
It is found that that nonzero $C_{z}(l)$ mainly distributes near
the interface and the integrated Hall conductance $\sigma_{\mathrm{H}}=e^{2}/2h$
\citep{Lu2021half,Gu2021spectral}. To facilitate the numerical calculation,
the chemical potential $\mu$ is set to deviate slightly from zero
to avoid the degeneracy at zero energy due to the gapless bottom surface.
Then we consider how the surface Hall conductance can be related to
measurable physical quantities. The Hall conductance is evaluated
with the periodic boundary condition in $xy$ plane. When we further
impose the open boundary condition in the $x$ direction, the quantum
confinement effect enforces the surfaces states into a series of subbands.
We plot the local density of state of the top left hinge labeled by
the star as shown in Fig. 3(c). The up moving states show heavier
spectral weight than the down moving modes. Although all these modes
are extended in the gapless region, they exhibit the chiral nature
and carry a circulating hinge current from Eq. (\ref{eq:current}).
As shown in Fig. 3(d), we plot the surface current distribution on
the perimeter of the semi-magnetic topological insulator bar for several
Zeeman fields with fixed chemical potential. In the gapless region,
both the current oscillation and asymptotic behavior $\sim|x+L/2|^{-3/2}$
of the envelope function can be well fitted by Eq. (\ref{eq:current})
(indicated by the red lines). While in the gapped region, the current
decays exponentially $\propto\exp(-\frac{2B}{\hbar v_{F}}|x+L/2|)$
from the interface.Thus, the half quantized Hall conductance can be
associated with the appearance of circulating currents (no hinge modes)
around the hinges. In the equilibrium case with a constant chemical
potential, the total current integrated over the width of the sample
is zero since the counter propagating currents localized on the opposite
edge cancel each other. The description of the equilibrium circulating
current that changes their magnitude as a function of the chemical
potential provides a useful framework for exploring the non-equilibrium
phenomena. As shown in Fig. 3(a), in the presence of the external
voltage bias $U$ between two side surfaces, each branch of the confinement
states generally has a different position-dependent chemical potential
and non-equilibrium carrier distribution. There is a net drop $eU$
in the chemical potential between two side surface states. From Eq.
(\ref{eq:current}), the edge current is proportional to the chemical
potential, the counter propagating edge currents can not compensate
and the total current becomes finite due to the voltage bias. The
spatial distribution of the edge current on the side surface is consist
with the layer-resolved Hall conductance, thus can be viewed as its
measurable physical quantity.

\paragraph{Discussion}

At last, we give a brief discussion on the quantum anomalous Hall
state and ``axion'' state which are realized in a topological insulator
thin film slab with parallel and antiparallel magnetic layer to the
top and bottom surface, respectively\citep{Mogi2017amagnetic,mogi2017tailoring,Xiao2018realization,Zhang2019topological,Liu2020robust}.
The former understanding is based on the Chern number calculation
with periodic boundary condition and the total Hall conductance counts
the the top and bottom surface together, yielding a $(1/2+1/2)$ or
$(1/2-1/2)$ quantized value, respectively. However, this picture
completely ignores the crucial effect from the lateral surface states
when the open boundary conditions are imposed. The proposed surface
states unfolding analysis can also be applied to these two cases with
both gapped top and bottom surfaces. In this situation, the gapless
side surface is sandwiched between the two gapped regions with same
or opposite mass depending on the relative magnetic orientation of
the two surfaces. Therefore, the side surface states for axion insulator
shares the same solutions for the parity anomalous semimetal. For
quantum anomalous Hall state, the eigenvalues for side surface states
are found to be $\varepsilon_{n}^{\mathrm{QAH}}(k_{y})=\hbar v_{F}\sqrt{k_{y}^{2}+[\frac{\pi}{L}n]^{2}}$
for $n\ge1$ and $\varepsilon_{n}^{\mathrm{QAH}}(k_{y})=\hbar v_{F}k_{y}$
for $n=0$ \citep{Note-on-SM}. The chiral state $n=0$ distributes
uniformly in the gapless side surface thus can be viewed as the generalized
Jackiw--Rebbi solution for mass domain wall \citep{Jackiw1976solitions}.
Since the chiral modes always coexist with unchiral modes for finite
chemical potential, it is found that the edge current shares the same
expression for parity anomalous semimetal (Eq. (\ref{eq:current}))
except for the reverse of the current at one interface. This result
implies that the power-law decaying edge current is a local property
at the interface between gapped and gapless regions that the Hall
conductances differ by one-half quantum. As seen from analytic exact
solutions, integer-quantized for the quantum anomalous Hall state
and zero Hall plateau for ``axion'' state can be understood in a
unified and comprehensive way from the aspect of the parity anomalous
semimetal.

J.Y.Z. and B.F. contributed equally to this work. This work was supported
by the National Key R\&D Program of China under Grant No. 2019YFA0308603
and the Research Grants Council, University Grants Committee, Hong
Kong under Grant Nos. C7012-21G and 17301220.

\end{document}